\pdfoutput=1

\documentclass[11pt]{article}
\usepackage{moriond}
\usepackage{graphicx}
\def\lesssim{\mathrel{\mathpalette\vereq<}}

\makeatletter
\def\vereq#1#2{\lower3pt\vbox{\baselineskip1.5pt \lineskip1.5pt
\ialign{$\m@th#1\hfill##\hfil$\crcr#2\crcr\sim\crcr}}}
\makeatother

\def\be{\begin{equation}}
\def\ee{\end{equation}}
\def\bea{\begin{eqnarray}}
\def\eea{\end{eqnarray}}

\begin{document}
\vspace*{4cm}
\title{SUPERSYMMETRY BREAKING MADE EASY, VIABLE, AND
  GENERIC\footnote{This talk is based on the collaboration with Yasunori
    Nomura.\cite{Murayama:2006yf,Murayama:2007fe}}} 

\author{HITOSHI MURAYAMA}

\address{Department of Physics, University of California, Berkeley, CA
94720\\
Theoretical Physics Group, Lawrence Berkeley National Laboratory,
Berkeley, CA 94720}

\maketitle\abstracts{The kind of supersymmetry that can be discovered
  at the LHC must be very much flavor-blind, which used to require
  very special intelligently designed models of supersymmetry
  breaking.  This led to the pessimism for some in the community that
  it is not likely for the LHC to discover supersymmetry.  I point
  out that this is not so, because a garden-variety supersymmetric
  theories actually can do this job.}

\section{Introduction}

LHC is coming!  It is finally taking us to the energy scale of the
weak interaction, $G_F^{-1/2} \approx 300$~GeV, known as an important
energy scale for more than seven decades since Fermi's 1933 paper on
the nuclear beta decay.  It is a historic moment in science and I am
very excited to be part of this new era.  Whenever physicists had
crossed a threshold of studying a new force, it resulted in a big
paradigm change.  The atomic scale (scale of quantum electrodynamics)
led to the revolutionary discovery of quantum mechanics.  The nuclear
scale (scale of quantum chromodynamics) revealed a new layer of matter
and showed the non-perturbative quantum field theory to be essential
in our description of nature.  We are all looking forward to whatever
paradigm change the weak scale will bring us.

However, there has been a growing concern in the community, especially
among the theorists, that we may not find anything surprising at the
electroweak scale.  I have been bitten by this bug, too.  The
reasoning is very simple.  If there is new rich physics below the TeV
scale such as supersymmetry and/or extra dimensions, why haven't we
seen its impact already on precision electroweak and flavor-physics
experiments?  Because we haven't seen such impacts, it is {\it
  unlikely}\/ that there is rich new physics below the TeV scale and
most likely {\it we will not find anything spectacular at the LHC.}\/

Even though I had plunged into this pessimism myself, I have now
completely turned around back to optimism.  I do think it is {\it
  quite likely for the LHC to find something exciting}\/ such as
supersymmetry.  I would like to tell you why I made this 180 degrees
change in my attitude.

The issue is the following.  Supersymmetry, if present at the TeV
scale, must be a broken symmetry because we have not seen any
superpartners yet.  The problem is that it has been believed that
breaking supersymmetry is very difficult, and certainly is not generic
among supersymmetric theories.  Even among those that do break
supersymmetry, they are rather difficult to use for constructing
phenomenologically viable models, hence most of them are dead on
arrival.  A very small fraction of the minority then survive, after an
elaborate model-building gymnastic, namely the ``alive'' theories are
``pockets of insurgency'' in the barren land.  An elaborate model is
like a beautiful artwork, {\it intelligently designed}\/, which is
what makes its creator(s) proud, but is by definition special and
fragile.  I can't stop feeling that Nature is unlikely to rely on
such a fragile elaborately built artwork for the foundation of its
inner working.

\begin{figure}[t]
  \centering
  \includegraphics[width=0.45\textwidth]{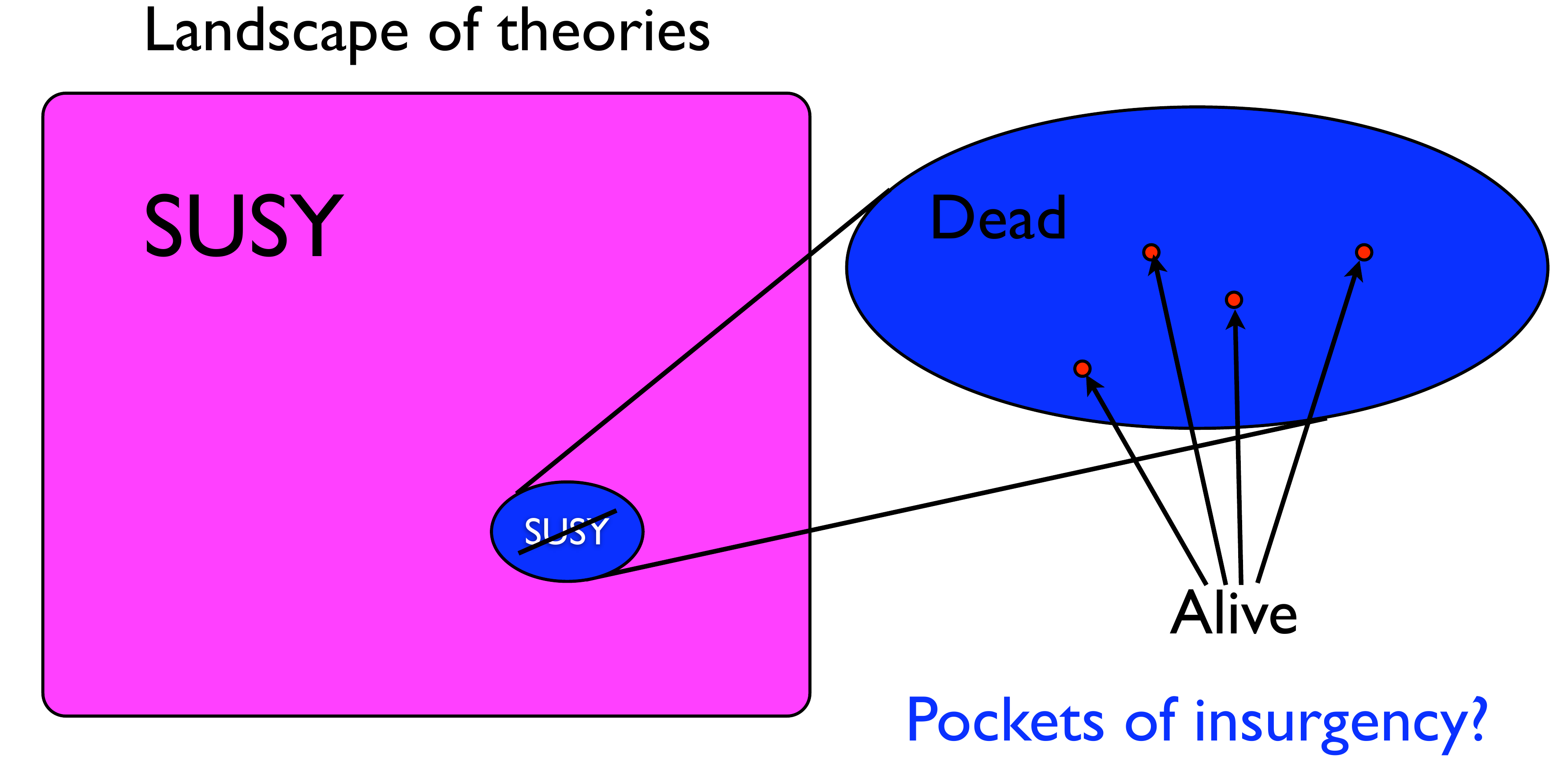} \qquad
  \includegraphics[width=0.45\textwidth]{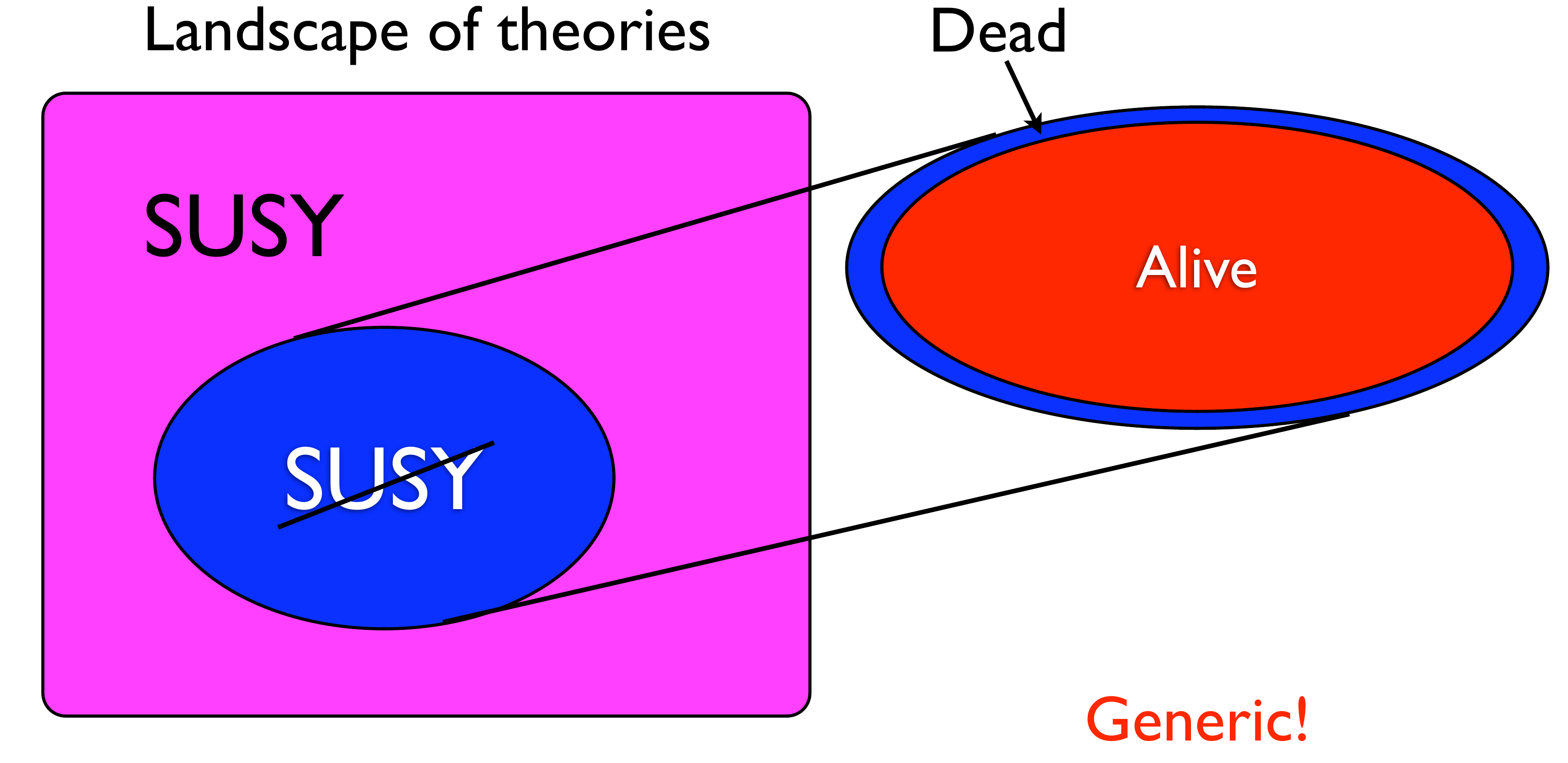}
  \caption{The ``landscape'' of supersymmetric theories.  (Left) I
    used to believe theories that break supersymmetry are very
    special, and much of them do not lead to phenomenologically viable
    theories.  ``Alive'' theories are very small minorities.  (Right)
    Now I believe there is a large class of theories that break
    supersymmetry, and a major fraction of them can be
    phenomenologically viable.  Hence, easy, viable, and generic.}
  \label{fig:landscape}
\end{figure}

I now feel the situation is very different from the previous
perception.  There is a large fraction of supersymmetric models that
can successfully break supersymmetry in a phenomenologically viable
fashion.  And it appears rather robust, namely a change in parameters,
such as choice of the gauge groups, number of flavors, does not spoil
its success.  This observation makes it much more plausible that
supersymmetry has a broad and robust foundation to be realistic, and
makes me feel that it could well be there waiting for us at the LHC.

\section{Hierarchy Problem}

It is often said that the hierarchy problem has been overemphasized as
a reason to expect rich physics at the LHC.  There is some truth in
it, because, after all, it is an aesthetic problem.  However, I'd like
to first remind you that the hierarchy problem was actually solved
once in the history of physics, and we can draw some lesson from
it.\cite{INS}

In the classical electromagnetism, the only dynamical degrees of 
freedom are electrons, electric fields, and magnetic fields.  When an 
electron is present in the vacuum, there is a Coulomb electric field 
around it, which has the energy of
\begin{equation}
    \Delta E_{\rm Coulomb} = \frac{1}{4\pi \varepsilon_{0}}\frac{e^{2}}{r_{e}}.
    \label{eq:ECoulomb}
\end{equation}
Here, $r_{e}$ is the ``size'' of the electron introduced to cutoff the
divergent Coulomb self-energy.  Since this Coulomb self-energy is
there for every electron, it has to be considered to be a part of the
electron rest energy.  Therefore, the mass of the electron receives an
additional contribution due to the Coulomb self-energy:
\begin{equation}
    (m_{e} c^{2})_{\it obs} = (m_{e}c^{2})_{\it bare} + \Delta E_{\rm 
    Coulomb}.
    \label{eq:self}
\end{equation}
Experimentally, we know (now) that the ``size'' of the electron is
small, $r_{e} \lesssim 10^{-17}$~cm.  This implies that the
self-energy $\Delta E$ is at least a few GeV, and hence the ``bare''
electron mass must be negative to obtain the observed mass of the
electron, with a fine cancellation like\footnote{Do you recognize
  $\pi$?}
\begin{equation}
    0.000511 = (-3.141082 + 3.141593)~{\rm GeV}.
\end{equation}
Even setting a conceptual problem with a negative mass electron aside,
such a fine cancellation between the ``bare'' mass of the electron and
the Coulomb self-energy appears troublesome.  In order for such a
cancellation to be absent, Landau and Lifshitz\cite{LL} concluded that
the classical electromagnetism cannot be applied to distance scales
shorter than $e^{2}/(4\pi \varepsilon_{0} m_{e} c^{2}) = 2.8\times
10^{-13}$~cm.  This is a long distance in the present-day particle
physics' standard.

\begin{figure}[t]
  \centerline{
    \includegraphics[scale=0.5]{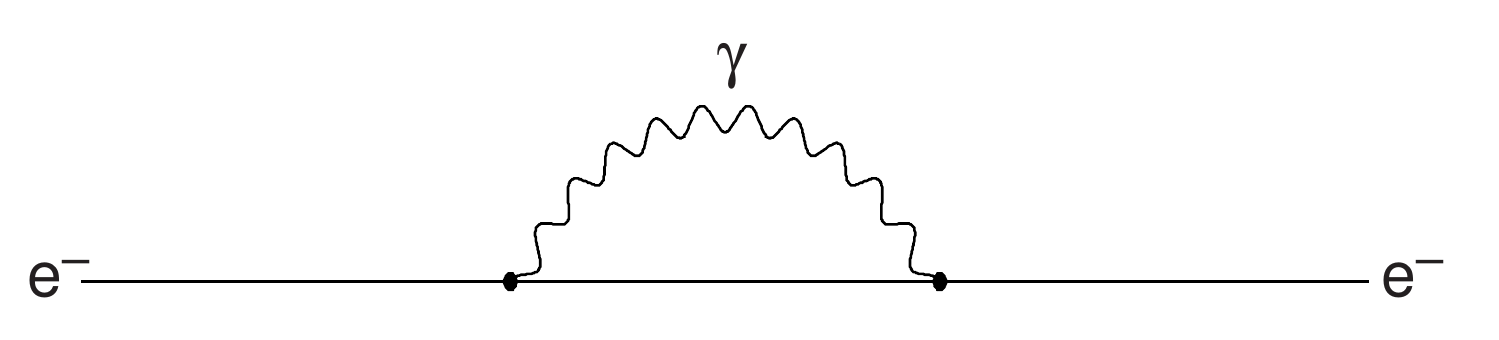}
    \includegraphics[scale=0.4]{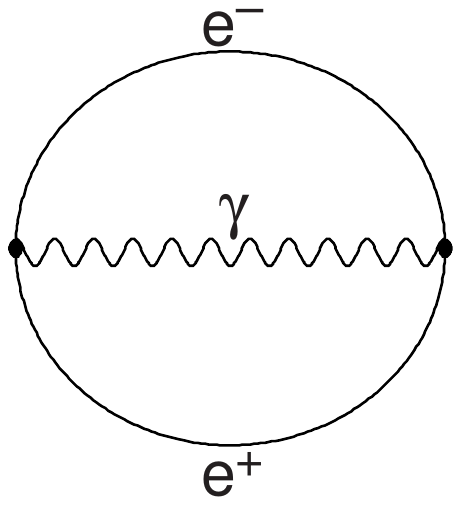}
    \includegraphics[scale=0.45]{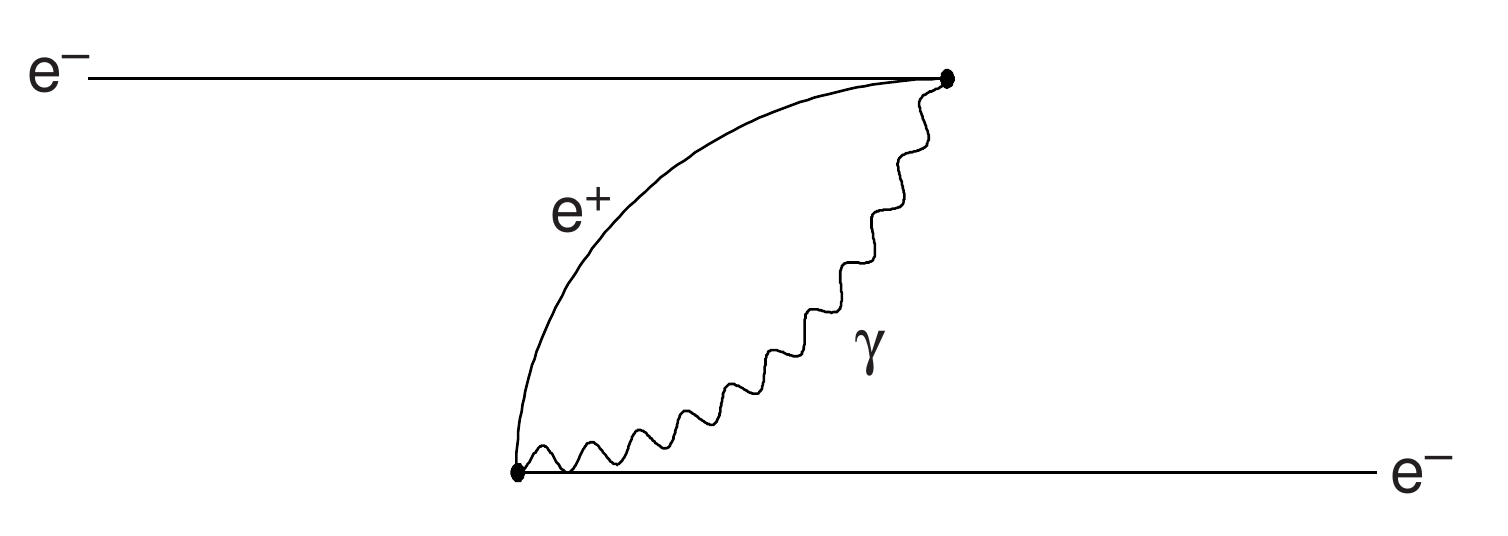}
  }
  \caption{(Left) The Coulomb self-energy of the electron.  (Middle)
    The bubble diagram which shows the fluctuation of the
    vacuum. (Right) Another contribution to the electron self-energy
    due to the fluctuation of the vacuum.}
  \label{fig:bubble}
\end{figure}

The resolution to this problem came from the discovery of the
anti-particle of the electron, the positron, or in other words by
doubling the degrees of freedom in the theory.  The Coulomb
self-energy discussed above can be depicted by a diagram
Fig.~\ref{fig:bubble}, left where the electron emits the Coulomb field
(a virtual photon) which is felt (absorbed) later by the electron
itself.  But now that we know that the positron exists, and we also
know that the world is quantum mechanical, one should think about the
fluctuation of the ``vacuum'' where a pair of an electron and a
positron appears out of nothing together with a photon, within the
time allowed by the energy-time uncertainty principle $\Delta t \sim
\hbar/\Delta E \sim \hbar/(2 m_{e} c^{2})$ (Fig.~\ref{fig:bubble},
middle).  This is a new phenomenon which didn't exist in the classical
electrodynamics, and modifies physics below the distance scale $d \sim
c \Delta t \sim \hbar c/(2 m_{e} c^{2}) = 200\times 10^{-13}$~cm.
Therefore, the classical electrodynamics indeed does hit its limit of
applicability at this distance scale, much earlier than $2.8 \times
10^{-13}$~cm as was exhibited by the problem of the fine cancellation
above.  Given this vacuum fluctuation process, one should also
consider a process where the electron sitting in the vacuum by chance
annihilates with the positron and the photon in the vacuum
fluctuation, and the electron which used to be a part of the
fluctuation remains instead as a real electron (Fig.~\ref{fig:bubble},
right).  V.~Weisskopf\cite{Weisskopf} calculated this contribution to
the electron self-energy, and found that it is negative and cancels
the leading piece in the Coulomb self-energy exactly:
\begin{equation}
    \Delta E_{\rm pair} = - \frac{1}{4\pi
      \varepsilon_{0}}\frac{e^{2}}{r_{e}} \ .
    \label{eq:Epair}
\end{equation}
After the linearly divergent piece $1/r_{e}$ is canceled, the 
leading contribution in the $r_{e} \rightarrow 0$ limit is given by
\begin{equation}
    \Delta E = \Delta E_{\rm Coulomb} + \Delta E_{\rm pair}
    = \frac{3\alpha}{4\pi} m_{e} c^{2} \log \frac{\hbar}{m_{e} c
      r_{e}}\ .
    \label{eq:DeltaE}
\end{equation}
There are two important things to be said about this formula.  First, 
the correction $\Delta E$ is proportional to the electron mass and 
hence the total mass is proportional to the ``bare'' mass of the 
electron,
\begin{equation}
    (m_{e} c^{2})_{\it obs} = (m_{e}c^{2})_{\it bare}
    \left[ 1 + \frac{3\alpha}{4\pi} \log \frac{\hbar}{m_{e} c r_{e} }
    \right].
    \label{eq:self2}
\end{equation}
Therefore, we are talking about the ``percentage'' of the correction, 
rather than a huge additive constant.  Second, the correction depends 
only logarithmically on the ``size'' of the electron.  As a result, 
the correction is only a 9\% increase in the mass even for an electron 
as small as the Planck distance $r_{e} = 1/M_{Pl} = 1.6 \times 
10^{-33}$~cm.  

The fact that the correction is proportional to the ``bare'' mass is a
consequence of a new symmetry present in the theory with the
antiparticle (the positron): the chiral symmetry.  In the limit of the
exact chiral symmetry, the electron is massless and the symmetry
protects the electron from acquiring a mass from self-energy
corrections.  The finite mass of the electron breaks the chiral
symmetry explicitly, and because the self-energy correction should
vanish in the chiral symmetric limit (zero mass electron), the
correction is proportional to the electron mass.  Therefore, the
doubling of the degrees of freedom and the cancellation of the power
divergences lead to a sensible theory of electromagnetism applicable
to very short distance scales.

In the Standard Model, the Higgs potential is given by
\begin{equation}
    V = m^{2} |H|^{2} + \lambda |H|^{4},
    \label{eq:V}
\end{equation}
where $v^{2} = \langle H \rangle^{2} = -m^{2}/2\lambda = (176~{\rm
GeV})^{2}$.  Because perturbative unitarity requires that $\lambda
\lesssim 1$, $-m^{2}$ is of the order of $(100~{\rm GeV})^{2}$. 
However, the mass squared parameter $m^{2}$ of the Higgs doublet
receives a quadratically divergent contribution from its self-energy
corrections.  For instance, the process where the Higgs doublets
splits into a pair of top quarks and come back to the Higgs boson
gives the self-energy correction
\begin{equation}
   \Delta m^{2}_{\rm top} = - 6 \frac{h_{t}^{2}}{4\pi^{2}} 
   \frac{1}{r_{H}^{2}}\ ,
    \label{eq:mu2top}
\end{equation}
where $r_{H}$ is the ``size'' of the Higgs boson, and $h_{t} \approx
1$ is the top quark Yukawa coupling.  Based on the same argument in
the previous section, this makes the Standard Model not applicable
below the distance scale of $10^{-17}$~cm, according to the
Landau--Lifshitz criterion.  This is the hierarchy problem.

The motivation for supersymmetry is to make the Standard Model
applicable to much shorter distances so that we can hope that the
answers to many of the puzzles in the Standard Model can be given by
physics at shorter distance scales.  In order to do so, supersymmetry
repeats what history did with the positron: doubling the degrees of
freedom with an explicitly broken new symmetry.  Then the top quark
would have a superpartner, the stop, whose loop diagram gives another
contribution to the Higgs boson self energy
\begin{equation}
   \Delta m^{2}_{\rm stop} = + 6 \frac{h_{t}^{2}}{4\pi^{2}} 
   \frac{1}{r_{H}^{2}}\ .
    \label{eq:mu2stop}
\end{equation}
The leading pieces in $1/r_{H}$ cancel between the top and stop 
contributions, and one obtains the correction to be
\begin{equation}
    \Delta m^{2}_{\rm top} + \Delta m^{2}_{\rm stop}
    = -6\frac{h_{t}^{2}}{4\pi^{2}} (m_{\tilde{t}}^{2} - m_{t}^{2})
    \log \frac{1}{r_{H}^{2} m_{\tilde{t}}^{2}}\ .
    \label{eq:Deltamu2}
\end{equation}

One important difference from the positron case, however, is that the
mass of the stop, $m_{\tilde{t}}$, is unknown.  In order for the
$\Delta m^{2}$ to be of the same order of magnitude as the tree-level
value $m^{2} = -2\lambda v^{2}$, we need $m_{\tilde{t}}^{2}$ to be not
too far above the weak scale.  TeV stop mass is already a fine tuning
at the level of a percent.  Similar arguments apply to masses of other
superpartners that couple directly to the Higgs doublet.  This is the
so-called naturalness constraint on the superparticle
masses.\cite{naturalness}

It is worth pondering if the mother nature may fine-tune.  Now that
the cosmological constant appears to be fine-tuned at the level of
$10^{-120}$, should we be worried really about the
fine-tuning\cite{Arkani-Hamed:2004fb} of $v^2/M_{Pl}^2 \approx
10^{-30}$?  In fact, some people argued that the hierarchy exists
because intelligent life cannot exist otherwise.\cite{Agrawal:1997gf}
On the other hand, a different way of varying the hierarchy does seem
to support stellar burning and life.\cite{Harnik:2006vj} We don't get
into this debate here, but we'd like to just point out that a
different fine-tuning problem in cosmology, horizon and flatness
problems, pointed to the theory of inflation, which in turn appears to
be empirically supported by data.  We just hope that proper solutions
will be found to both of these fine-tuning problems and we will see
their manifestations at the relevant energy scale, namely TeV.

\section{Why We Were Pessimistic}

Supersymmetry or not, we expect some interesting physics to appear
below TeV scale if the hierarchy problem is to be avoided by some
stabilization mechanism.  The problem is that it is difficult to
understand why we have not seen its impact on flavor-changing neutral
currents especially in the beautiful $B$ physics data and electroweak
precision measurements.  Are we on the wrong track to naively hope
that the stabilization mechanism is just around the corner?  Or is
there rather a good reason why it doesn't show its fingerprints
despite our best detective work?  This is a question that applies to
{\it any}\/ candidate physics beyond the standard model at the TeV
scale.

The problem with supersymmetry is well-known, having been discussed
already for a several decades.  In the Minimal Supersymmetric Standard
Model (MSSM), which is the supersymmetric extension of the Standard
Model with the smallest particle content, there are staggering 107
additional parameters beyond the nineteen in the Standard Model.  And
if you throw dice in this huge parameter space, you almost always end
up with a parameter set that is already excluded by the data.  For
example, the off-diagonal elements in the mass-squared matrices must
be less than a few per mill of the mass eigenvalues for three types of
squark and two types of slepton mass matrices.  Also the mixing
between the scalar partners of left- and right-handed fermions need to
be very much identical to the mixing among the fermions.  Overall, the
probability of ``hitting'' the phenomenologically viable parameter
sets would be down by a product of many factors of hundreds.  Why are
the unwanted parameters small, supersymmetry-breaking effects
flavor-blind?  Unless there is a good reason, the whole idea of
sub-TeV supersymmetry to stabilize the hierarchy appears a remote
chance.

In addition, breaking supersymmetry appears difficult and highly
non-generic among supersymmetric theories.  Known models require a
specific choice of gauge groups and matter content, often together
with a special choice of the superpotential terms with a global
symmetry imposed; global symmetries are usually regarded unlikely in a
fundamental theory of quantum gravity such as string theory.

There are several popular mechanisms to achieve flavor-blind
supersymmetry breaking: gauge mediation,\cite{AlvarezGaume:1981wy}
gaugino mediation,\cite{Kaplan:1999ac} and anomaly
mediation.\cite{Randall:1998uk} The supersymmetry-breaking effects are
``mediated'' to the supersymmetric standard model via gravity or gauge
interactions, guaranteeing their flavor-blindness.  Even though these
mechanisms do work, my problem has been that the models must be
written in a very careful and elaborate fashion.  Small changes in the
models, such as a different choice of the gauge group or matter
content, tend to destroy their success, such as restoring
supersymmetry, allowing for flavor-dependent effects, destabilizing
the vacuum.  My feeling has been that they do not represent a likely
choice by Nature.

\begin{figure}[tbhp]
  \centering
  \includegraphics[height=1.6in]{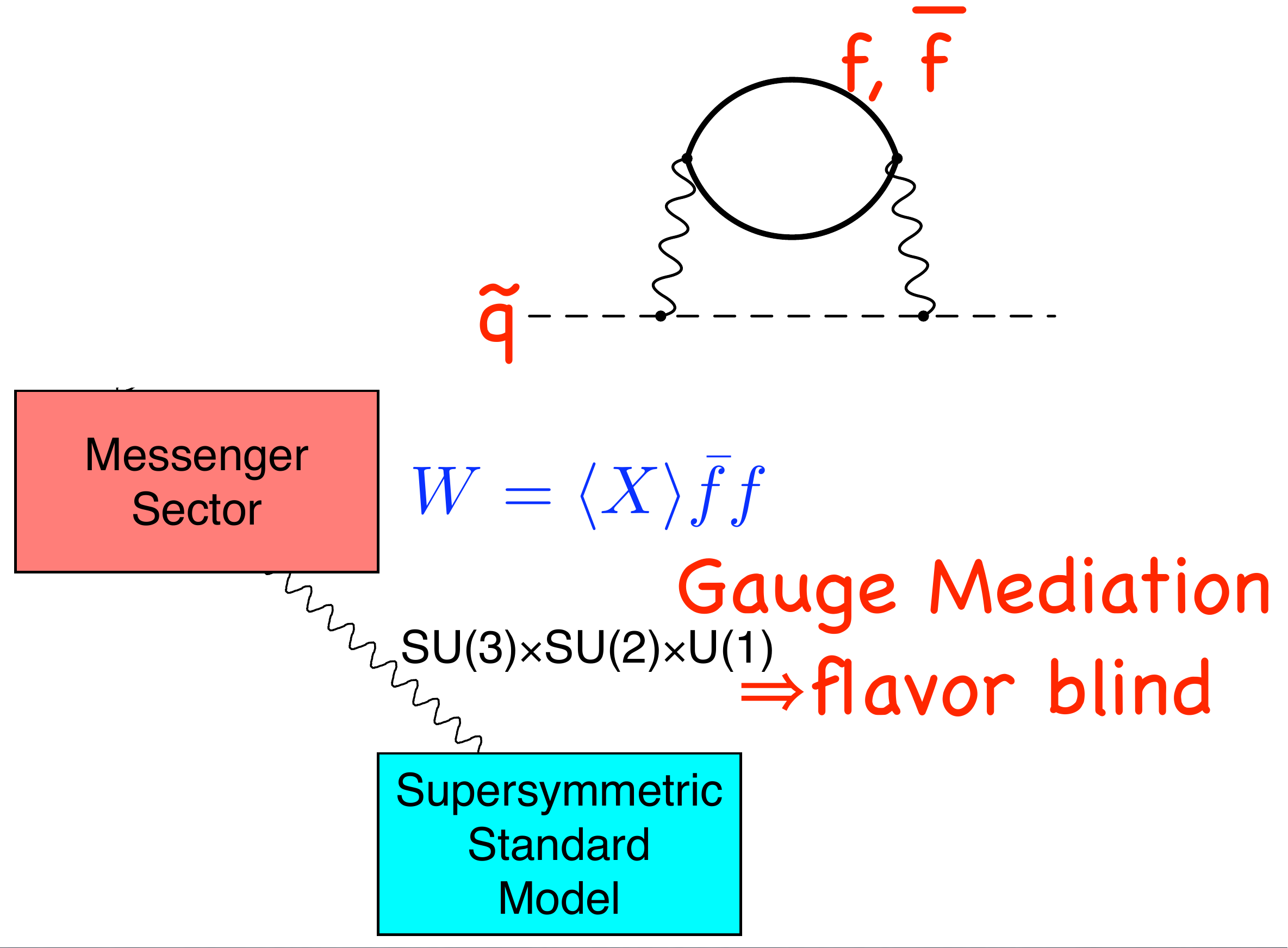} \qquad
  \includegraphics[height=1.6in]{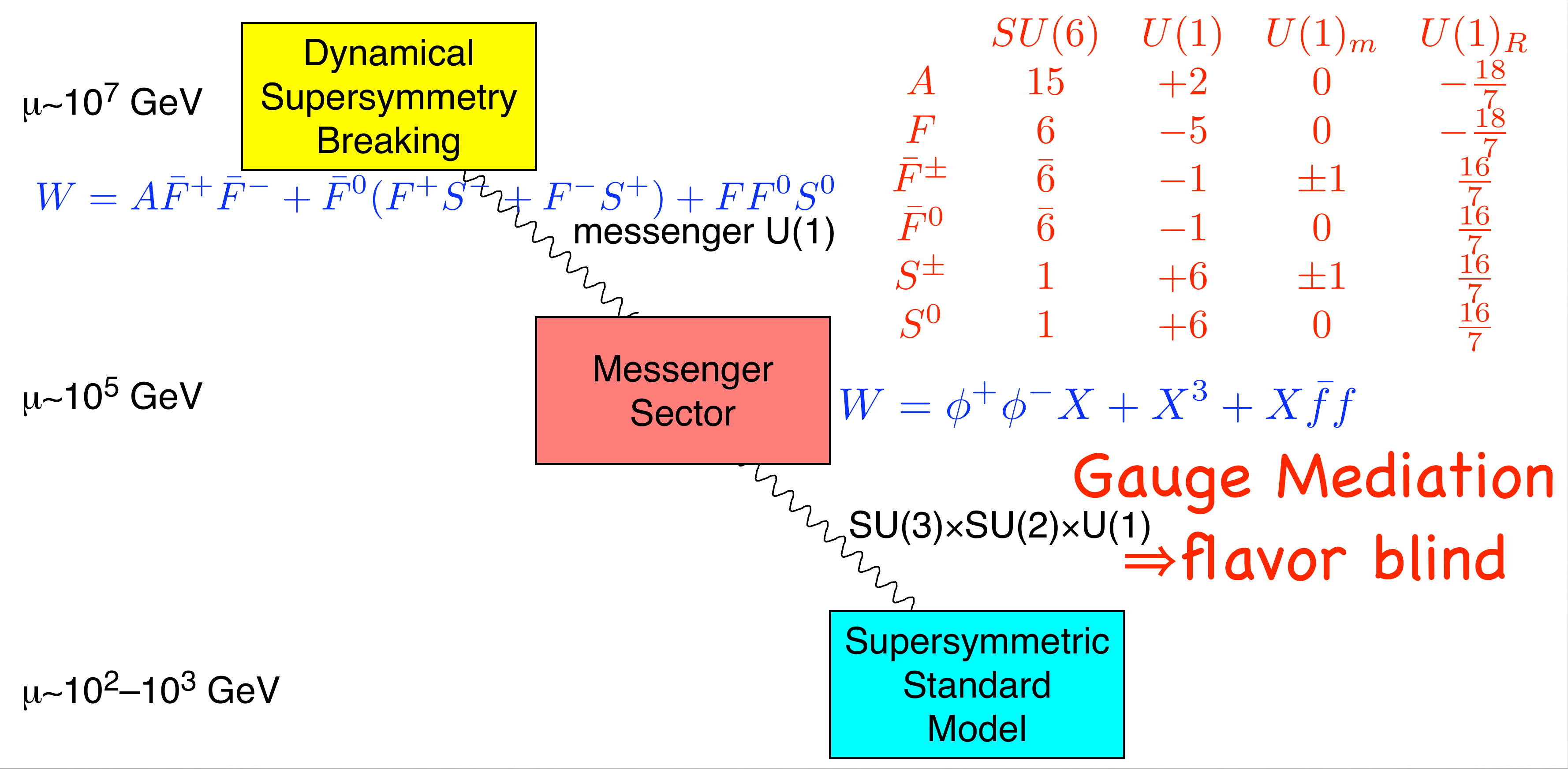}
  \caption{(Left) A beautiful mechanism for flavor-blind supersymmetry
    breaking (gauge mediation).  (Right) However, a full model is
    quite involved and elaborate.\protect\cite{Dine:1995ag}}
  \label{fig:gauge-mediation}
\end{figure}

Let us see how careful and elaborate models are needed with an
example.  Gauge mediation, at the first sight, is a beautiful idea.
It has a set of {\it messenger particles}\/ $f$ and $\bar{f}$ that
carry standard-model quantum numbers.  Once supersymmetry is broken by
a vacuum expectation value of field $X$, the messengers do not have a
supersymmetric spectrum; the masses of the bosons are split from those
of the fermions.  Then their loops induce masses for squarks,
sleptons, and gauginos (Fig.~\ref{fig:gauge-mediation}, left).  So far
so good.

However, one has to ask the question how the supersymmetry-breaking
expectation value is generated for $X$.  It requires a separate gauge
theory with a rather complicated particle content and specific
potential, on which a global symmetry is imposed to make sure that it
breaks supersymmetry.  Only a small fraction of supersymmetric models
serve this purpose.  This gauge theory is coupled to the messengers
also in a constrained specific way through yet another gauge
interaction and singlet particles (Fig.~\ref{fig:gauge-mediation},
right).  Overall, we need nearly decoupled three ``boxes'' to make the
whole mechanism possible.

If a realistic model of nature has to rely on a carefully constructed
elaborate mechanism, even though it is logically possible, I am not
sure if that is Mother Nature's choice.  In some sense, we rely on
supersymmetry to avoid fine-tuning in electroweak symmetry breaking
and flavor, but we end up {\it fine-picking}\/ or {\it intelligently
  designing}\/ a model.  Even though it is a philosophical point and
not very scientific, I would be very much happier if we can do
without {\it fine-picking}\/.  If we think Nature doesn't fine-tune,
she probably doesn't fine-pick either; she is way smarter than us,
after all.

\section{The New Generic Scheme}

My main point here is that I actually do not need to be very
intelligent to achieve flavor-blind supersymmetry breaking.  Pretty
much the dumbest supersymmetric extension of the standard model would
do it. 

We still need the messengers, non-chiral particles coupled to the
standard model.  Such particles are known to arise generically in
string theory, and people have been trying hard to get rid of such
{\it junks}.\/ In addition, again generically, one expects other gauge
groups, with their own quarks; more junks.  Most of them tend to be
non-chiral.  I claim these junks are precisely what we need.  No
fine-picking.\footnote{For instance, the first three-generation models
  from heterotic strings based on Tian-Yau manifold has six
  vector-like families, and an extra $E_8$ for disposal.  The only
  significant requirement here is that there should better not be
  chiral particles that couple to both the standard model and the
  other gauge groups.  I thank Mirjam Cveti\v{c} on this point.}

\begin{figure}[tbhp]
  \centering
  \includegraphics[height=1.6in]{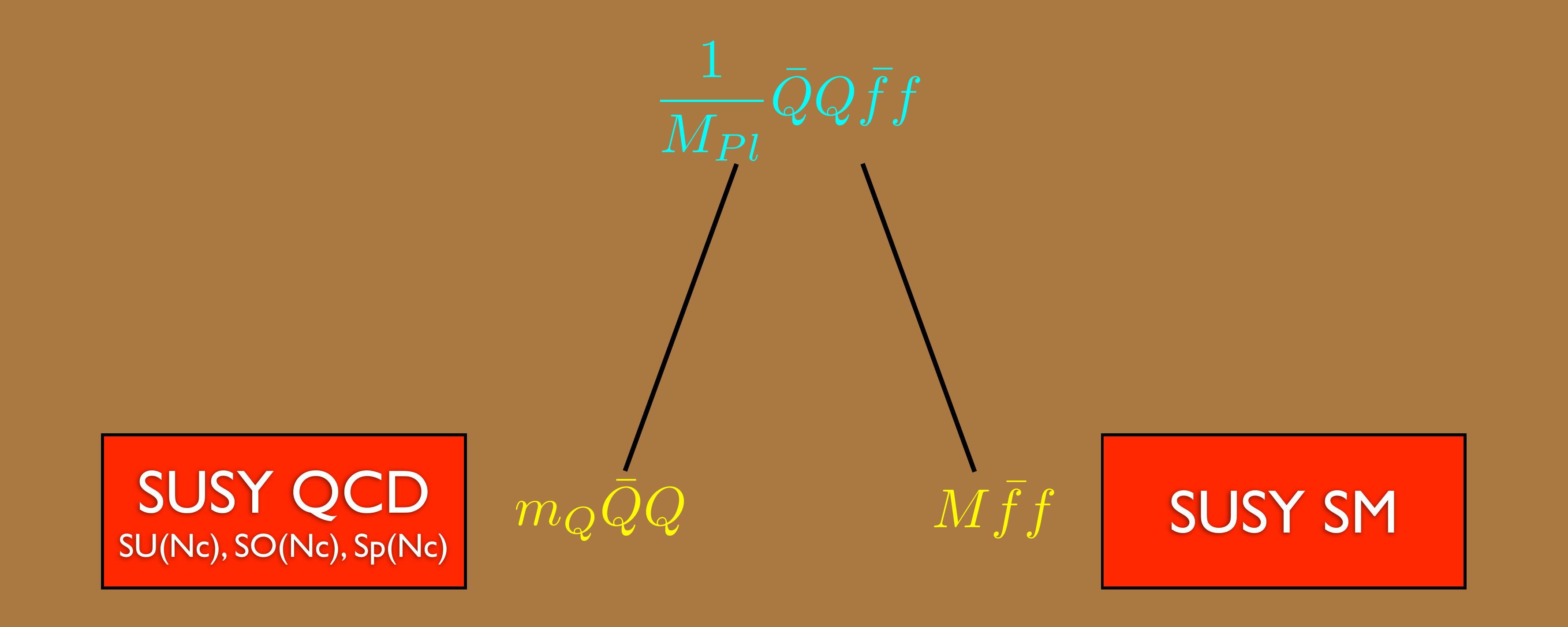}
  \caption{Generic full models of flavor-blind supersymmetry
    breaking.\protect\cite{Murayama:2006yf} }
  \label{fig:MY}
\end{figure}

We do not impose a global symmetry on the model.  We write the most
general potential consistent with the gauge symmetries.  The lowest
dimension operator that couple the messengers and the other quarks is
precisely what brings the supersymmetry-breaking effects to the
messengers, and hence to the standard model in a flavor-blind way.
The gauge groups can be pretty much anything, $SU(N_c)$, $SO(N_c)$,
$Sp(N_c)$; all classical groups.  (Let us focus on $SU(N_c)$ case below
but the other groups do the same thing.)  The parameters ({\it
  e.g.}\/, the gauge couplings, mass scales, etc) do not need to be
tuned against each other.  They have to satisfy certain inequalities,
such that the possibly flavor-sensitive Planck-scale physics would be
subdominant compared to the gauge-mediated contributions, or the
number of flavors in the other ``QCD'' is in the range $N_c < N_f <
\frac{3}{2} N_c$.  No fine-tuning here.

How this simple scheme works requires a little technical discussion.
For this range of the $N_f$, the other ``QCD'' becomes strong at some
energy scale $\Lambda$, and the low-energy limit is known to be
described by yet another gauge theory $SU(N_f - N_c)$, whose quarks
$q$ and $\bar{q}$ couple to the mesons $S_{ij} = \bar{Q}_i Q_j$ $(i,j
= 1, \cdots, N_f)$ of the original ``QCD'' through the potential
\begin{equation}
  W = m_Q^{ij} \Lambda S_{ij} - S_{ij} \bar{q}^i q^j.
\end{equation}
We assume $m_Q^{ij} \ll \Lambda$.  Then this potential does not have a
solution to the supersymmetric minimum
\begin{equation}
  \frac{\partial W}{\partial S_{ij}} = m_Q^{ij} \Lambda - \bar{q}^i
  q^j = 0
\end{equation}
because $m_Q^{ij}$ has rank $N_f$ while $\bar{q}^i q^j$ rank
$N_f-N_c$.  This supersymmetry-breaking minimum is actually a local
minimum,\cite{Intriligator:2006dd} but the tunneling to the global
supersymmetric minimum has an exponentially long lifetime.  The lowest
dimension operator for the messengers become
\begin{equation}
  M \bar{f} f + \frac{\Lambda}{M_{Pl}} \langle S \rangle \bar{f} f,
\end{equation}
exactly what is needed for the gauge mediation
(Fig.~\ref{fig:gauge-mediation}, left).

In fact, any models that break supersymmetry can be used the same
way.\cite{Murayama:2007fe} Just the vector-like ``junks'' coupled to
the standard model, and the lowest dimension operator to link them.

This, I believe, is a good news for string theory.  String theory is
now believed to have many many solutions, some $10^{500}$ of them.
The vast majority of them have huge cosmological constants and do not
resemble our universe; they do not support life and do not get
observed by scientists like us.  We have lost very many solutions by
this cut.  Getting standard model is another severe cut on the
numbers.  It would be nice if a large fraction of the remaining
solutions would lead to successful supersymmetry breaking and
phenomenologically viable models; then Nature may well have given us
one.  The simple scheme presented here suggests that a significant
fraction of the remaining solutions indeed may well do so.

Experimental consequences are pretty much the same as the
phenomenology of gauge-mediated models people have been discussing in
the literature.  The dark matter particle is the gravitino.  Even
though it is a bad news for direct detection experiments, it opens up
an interesting possibility of producing gravitinos of spin 3/2 at
colliders.  There may be extra photons or long-lived charged particles
in the supersymmetry events.  The mass spectrum of superparticles tell
us about the quantum numbers of the messengers even though they are
beyond the reach of direct production.  Finally, the linear-collider
precision of superparticles may reveal the presence of the light
particles in the other ``QCD''.\cite{preparation}

\section{Conclusions}

Even though supersymmetry is a beautiful idea to solve the fine-tuning
or hierarchy problem, we will not see it at the LHC unless it comes
out flavor-blind.  Theorists used to be a kind of control freak to
write special models that ensure the flavor-blindness of
supersymmetry.  This {\it fine-picking}\/ of models made us
uncomfortable, feeling the chance for its discovery at the LHC remote.
But after some more thoughts, it turned out that we don't really need
fine-picking to break supersymmetry in a flavor-blind fashion.  It is
{\it easy}\/ to write a model, which is phenomenologically {\it
  viable}\/, and the scheme is very {\it generic}\/, a kind of
spectrum one expects from the string theory.  Pretty much the dumbest
extension of the supersymmetric standard model would do the job.

I do not share anymore the spreading concern that LHC is not likely
to discover exciting physics because we have not seen any hints of it
yet.  Quite generically in the ``landscape'' of supersymmetric
theories, we expect the superparticles to come out flavor-blind and
therefore well hidden from the current beautiful data.  I suspect
this is probably not specific to supersymmetry.  More thoughts may
well reveal why we have not seen hints of TeV-scale new physics yet,
even though it is waiting to be discovered at the LHC.

\section*{Acknowledgments}

This work was supported by the U.S. DOE Contract DE-AC03-76SF00098 and
the NSF grant PHY-04-57315.  The work of Y.N. was also supported by
the NSF grant PHY-0555661, by a DOE OJI award, and by an Alfred
P. Sloan Research Fellowship.

\end{document}